\documentclass[%
 reprint,
superscriptaddress,
 amsmath,amssymb,
 aps,
pra,
]{revtex4-2}

\usepackage{graphicx}
\usepackage{dcolumn}
\usepackage{bm}

\begin{document}


\title{Impurity-induced trimer formation across the BCS-BEC 
crossover}

\author{M. C. Gordillo}
\email{cgorbar@upo.es}
\affiliation{Departamento de Sistemas F\'{\i}sicos, Qu\'{\i}micos y Naturales, Universidad Pablo de
Olavide, Av.  Rectora Rosario Valpuesta, 1, E-41089 Dos Hermanas, Seville, Spain}
\affiliation{Instituto Carlos I de F\'{\i}sica Te\' orica y Computacional, Universidad de Granada, E-18071 Granada, Spain.}

\author{J. Boronat}
\affiliation{Departament de F\'{\i}sica,
Universitat Polit\`ecnica de Catalunya,
Campus Nord B4-B5, 08034 Barcelona, Spain}

\date{\today}

\begin{abstract}
By means of diffusion Monte Carlo calculations, we study the behavior of a 
single impurity embedded in two balanced Fermi baths across the BCS-BEC 
crossover. The impurity interacts attractively with both baths through 
P\"oschl-Teller potentials of different strengths. That interaction is such that does not support an impurity--bath
two-body bound state. When that attraction is sufficiently strong, 
we observe the formation of trimers composed of the impurity and one atom from 
each bath. These clusters are Borromean on the BCS side, where no two-body 
subsystem is bound, and evolve continuously into tango-like trimers on the BEC 
side. Subtracting from the total energy the binding energy of one trimer and 
that of the remaining bath-bath dimers restores the equation of state of the 
impurity-free system, supporting a dilute-cluster interpretation. The analysis 
of interparticle distances shows that the embedded trimer is not simply a dimer 
dressed by the impurity, but a genuine three-body cluster formed inside the 
two Fermi baths.

\end{abstract}

\maketitle

\section{Introduction}

Ultracold atomic gases are an interesting field of study both in their own 
right
and as models for the behavior of more complex systems \cite{Bloch}. In 
particular, 
the possibility of tuning their interactions, typically by means of an external 
magnetic field, makes a wide range of physical phenomena experimentally 
accessible in regimes where the relevant parameters can be accurately controlled 
\cite{resonances}. One of the most remarkable examples of this control is the 
experimental realization of the Bardeen-Cooper-Schrieffer (BCS) to Bose-Einstein 
condensate (BEC) crossover in two-component Fermi gases \cite{bec1,bec2,bec3}. 
In these experiments, the
interaction strength was continuously changed,  connecting a regime of  
Cooper 
pairs (BCS) to another one  made of dimers formed by one atom from 
each species (BEC). This crossover, produced via a broad Feshbach resonance, 
has also been successfully described using quantum Monte Carlo calculations 
\cite{panda,jordibec,AFQMC} by changing a single parameter, the s-scattering 
length of the two species involved. 

The same experimental control also allows studying  mixtures with three 
distinguishable 
fermionic components. In particular, the lowest three hyperfine states of $^6$Li 
provide a natural realization of such systems, with tunable interactions between 
pairs of different states
\cite{li6-0,li6-1,li6-2,Braaten2010,Nakajima2010,Science}. These mixtures are 
especially 
interesting because, in addition to pairing, they support the formation of 
trimers with three different atoms \cite{Braaten2006,Naidon}. As a first 
approximation to the full three-component mixture, in this work we study a 
single impurity embedded in two interacting Fermi baths. This setup differs from 
the standard problem of a single particle (polaron) interacting with a 
many-body background, which has been widely investigated in previous literature 
(see for instance Ref. \cite{Massignan2014} for a review).  The main difference 
is the possibility of three-body correlations that  may lead to the formation 
of trimers. 

This last possibility has been rarely addressed in the literature. A notable 
exception is 
the work of Nishida \cite{Nishida} that uses a variational approach to deal with 
a single impurity interacting with two baths around a narrow Feshbach resonance, 
characterized by the s-wave scattering length and the resonance range. In that work, all the pairs share the same 
interactions and they find a transition from atom to trimer under
conditions different from the ones considered in this work.
Yi and Cui \cite{cui} considered an impurity immersed in a two-component Fermi superfluid across the BCS-BEC crossover using a variational ansatz built on the BCS state. They found a strong coupling between impurity-like and trimer-like configurations, manifested as avoided level crossings whose widths increase towards the BEC side.
The same happens in Ref. \onlinecite{2baths}, in which a single 
impurity immersed in two non-interacting baths undergoes a transition between a 
polaron and a trimer.  Finally, a perturbative treatment of
the weakly coupled polaronic branch in Ref.~\onlinecite{Pierce2019} suggests 
the 
existence of trimers in proper conditions. 

In the present work, 
and in contrast with previous approaches, the two interacting Fermi baths and 
the 
impurity are treated explicitly within a microscopic Hamiltonian, without 
assuming a mean-field description of the background or restricting the 
impurity-bath interaction to the perturbative regime. By varying the interaction 
between the bath components across the BCS-BEC crossover, we 
study the competition between bath pairing and impurity-induced three-body 
correlations. This allows us to follow the formation and evolution of bound 
trimer states inside the medium, from the BCS regime, where no two-body bound 
states are present, to the BEC regime, where the bath fermions form dimers.

\section{Method}

We considered a system formed by a single impurity and two fermionic baths, labelled
as species 1 and 2.  Both baths include the same number of atoms, $N_1=N_2=N$, all of them with the same mass $m$, mass that is also the one of the single impurity. 
With all that in mind,  we have chosen to model our system by the 
Hamiltonian
\begin{equation}
H =
-\frac{\hbar^2}{2m}\sum_{i=1}^{N_T}\nabla_i^2
+
\sum_{i=1}^{N_1}\sum_{j=1}^{N_2}
V_{12}(r_{ij})
+
\sum_{\alpha=1}^2 \sum_{i=1}^{N_{\alpha}}
V_{{\alpha}3}(r_{i3}),
\label{Hamiltonian} 
\end{equation}
with $N_T=N_1+N_2+1$ the total number of atoms, $3$ being the label of the impurity. 
Here, $r_{ij}=|\mathbf{r}_i-\mathbf{r}_j|$ is the distance between particles
belonging to different baths, and
$r_{i3}=|\mathbf{r}_i-\mathbf{r}_3|$ is the distance between the impurity and 
any atom of the other two species.  
As the density of the gas is extremely small, we neglect intraparticle 
interactions and consider only atom-atom interactions
 between particles of baths $1$ and $2$,  denoted as $V_{12}$,  
The
impurity-bath potential is the same for both species,
$V_{13}(r_{i3})=V_{23}(r_{j3})$.  All two-body interactions are described by an
attractive P\"oschl-Teller potential of the form \cite{panda,LiK}
\begin{equation}
V_{\alpha\beta}(r)
=
-\frac{\hbar^2}{m_r} \frac{V_{\alpha\beta}^{'}}{\cosh^2\left(\mu_{\alpha\beta} r\right)} ,
\label{pt}
\end{equation}
where $V_{\alpha\beta}^{'}>0$ determines the interaction strength and
$\mu_{\alpha\beta}^{-1}$ defines its range. $m_r$ is the reduced mass of all 
particle pairs, in this case $m_r$ = $m$/2.  In this work, we varied the value 
of $V_{12}$ to go from BCS to BEC regimes, i.e., we choose values of 
the potential well that, while attractive,  do not support a two-body bound 
state (BCS), or are deep enough to produce a single one (BEC). On the other 
hand, the $V_{13}^{'}=V_{23}^{'}$ parameter was always tuned to be shallow enough not to 
produce impurity-atom dimers.   There is a one-to-one relationship between 
$V_{\alpha\beta}^{'}$ and the corresponding scattering length $a_{\alpha\beta}$ of 
the P\"oschl-Teller potential given by
\begin{equation}
a_{\alpha\beta}=
\frac{1}{\mu_{\alpha\beta}}
\left[
\frac{\pi}{2}\cot\left(\frac{\pi\lambda}{2}\right)
+\gamma
+\psi(\lambda)
\right]
\label{sca}
\end{equation}
with 
\begin{equation}
\lambda=
\frac{1}{2}
\left(
1+\sqrt{1+\frac{8V_{\alpha\beta}^{'}}{\mu_{\alpha\beta}^2}}
\right),
\label{lambda}
\end{equation}
Therefore, we use interchangeably potential strengths,
$V_{\alpha\beta}^{'}$, and scattering lengths,
$a_{\alpha\beta}$, to characterize the interactions.
Here, $\psi(x)$ is the digamma function \cite{abramowitz,dlmf}, and $\gamma$ = $-\psi(1)$ is the Euler-Mascheroni constant.  All distances and scattering lengths are given in units of $\mu_{13}^{-1}$.  We also fixed $\mu_{12} = 3\mu_{13}/4$ and
define $\epsilon_0$=$\hbar^2 \mu_{13}^2/m_r$ as our unit of energy, so the interaction between atoms can be written as:
\begin{equation}
V_{\alpha\beta}(r)
=
-\frac{V_{\alpha\beta}}{\cosh^2\left(\mu_{\alpha\beta} r\right)} ,
\label{pt2}
\end{equation}
with $V_{\alpha\beta} = V_{\alpha\beta}^{'} \epsilon_0/\mu_{13}^2$.  The corresponding unitary limits,
$a_{\alpha\beta} \rightarrow \infty$ for the two interactions being
$V_{12}=9\epsilon_0/16$ and $V_{13}=\epsilon_0$, respectively.  This means that in the 
$0.42\epsilon_0<V_{12}<0.85\epsilon_0$ range considered here, the potential (\ref{pt}) 
supports at most one bound state.  
On the other hand, $V_{13} \leq \epsilon_0$ in all cases, what implies negative scattering lengths.  

To solve the Schr\"odinger equation derived from the
Hamiltonian in Eq.  \ref{Hamiltonian}, we used the fixed-node diffusion
Monte Carlo (FN-DMC) method, that gives us an upper bound for 
the energy of the ground state of a system of fermions \cite{hammond} and goes beyond the mean-field, variational or pertubative approaches of the previous literature \cite{Nishida,cui,2baths,Pierce2019}.  The 
quality of the results is determined by the trial wave function used to guide 
the diffusion process, in particular to 
its nodal surface, which is not changed along the simulation to keep the sign 
problem under control.
In this work, 
we will use an extension of the wave functions used in Ref. 
\onlinecite{jordibec} that includes impurity-baths correlations. 
For the BCS region of the two baths interactions, we chose:
\begin{equation}
\begin{aligned}
\psi_{\rm JS}(\mathbf{R})
={}&
\prod_{\alpha=1}^{2}
\prod_{k=1}^{N_{\alpha}}
\varphi_{\alpha 3}(r_{k3})
\prod_{i=1}^{N_1}
\prod_{j=1}^{N_2}
\varphi_{12}(r_{ij})
\\
&\times
\left[ {\cal A}\prod_{i,\alpha} 
e^{i{\bf k}_\alpha\cdot{\bf r}_i} \right] \left[ {\cal 
A}\prod_{i^\prime,\alpha} 
e^{i{\bf k}_\alpha\cdot{\bf r}_{i^\prime}} \right]
&\times
\end{aligned}
\end{equation}
where each determinant contains a set of plane waves, $e^{i\mathbf{k}_a\cdot\mathbf{r}_i}$,  
defined by wave vectors
$\mathbf{k}_a=(2\pi/L)(l_{a,x},l_{a,y},l_{a,z})$, where $L$ is the side
length of the simulation cell, and $l_{a,x}$, $l_{a,y}$, and $l_{a,z}$ are
integers.  $a$ goes from 1 to $N_1$=$N_2$. Here, $\mathbf{R}= ({\bf r_1}, {\bf r_2}, \cdots, {\bf r_{N_T}})$
stands for the set of coordinates of all the atoms, 
 those belonging to the baths and the one of the impurity. 
The functions $\varphi_{12}(r)$ and
$\varphi_{13}(r)$=$\varphi_{23}(r)$ could naively be taken as the solutions
of the corresponding zero-energy scattering equations. Instead, we modify
them in order to avoid the numerical difficulties associated with their
typically long tails. To this end, we use the zero-energy scattering
functions up to a matching point $L_{\alpha\beta}<L/2$, and set
\[
\varphi_{\alpha\beta}(r)
=
A_{\alpha\beta}
+
B_{\alpha\beta}
\exp(-\delta_{\alpha\beta} r),
\]
for $r>L_{\alpha\beta}$. The matching points $L_{\alpha\beta}$ are optimized 
variationally, whereas
the coefficients $A_{\alpha\beta}$, $B_{\alpha\beta}$, and
$\delta_{\alpha\beta}$ are fixed by requiring continuity of the wave functions
and of their first and second derivatives at $r=L_{\alpha\beta}$.

On the other hand,  in the BEC regime of the two baths we chose as the trial 
function
\begin{equation}
\begin{aligned}
\psi_{\rm P}(\mathbf{R})
&=
{\cal A} \left( \phi(r_{11^\prime})\phi(r_{22^\prime})...\phi(r_{N_\uparrow 
N_\downarrow})\right)
\prod_{\alpha=1}^{2}
\prod_{k=1}^{N_{\alpha}}
\varphi_{\alpha 3}(r_{k3}) .
\end{aligned}
\end{equation}
Here, $\phi_{12}(r_{i,j})$ is the solution of the two-body Schr\"odinger equation for each pair $i,j$, 
when $V_{12}(r_{i,j})$ supports a single bound state. 
That solution can be written as proportional to:
\begin{equation}
\phi(r)=
\frac{\tanh\left(\mu_{12} r\right)}
{\cosh^{\nu}\left(\mu_{12} r\right)} \frac{1}{r} ,
\end{equation}
with
\begin{equation}
\nu =
-\frac{3}{2}
+
\frac{1}{2}
\sqrt{
1+\frac{8V_{12}^{'}}{\mu_{12}^{2}}
}\, ,
\end{equation}
by ignoring a normalization constant irrelevant in the FN-DMC algorithm.  The 
binding energy between two atoms, in that regime is
\begin{equation}
\epsilon_b
=
-\frac{\mu_{12}^2}{8}
\left(
-3+\sqrt{1+\frac{8V_{12}^{'}}{\mu_{12}^2}}
\right)^2
\label{binding}
\end{equation}
in units of $\epsilon_0$. In the unitary limit for that interaction, we have
$V_{12}^{'}$ = $\mu_{12}^2$,  which implies $\epsilon_b=$0 and $\nu$=0, 
with~\cite{panda}
\begin{equation}
\phi(r)=
\frac{\tanh\left(\mu_{12} r\right)}{r} \ . 
\end{equation}
In order to ensure that
the mean interparticle distance is much larger than the
range of the potential, we used $n(\mu_{13}^{-3}) = 2.48 \times 10^{-6}$,  with 
$n$ the number density. This corresponds to $N_T$ = 67 particles (both baths and the impurity) in a unit cell of length $L$ = 300$\mu_{13}^{-1}$,  the density corresponding to each bath being $n_B(\mu_{13}^{-3})= 1.22 \times 10^{-6}$.  This size of the simulation cell is a good balance between accurate sampling and diluteness.  From that value,  we can obtain the Fermi vector as:
\begin{equation}
 k_F = (6 \pi^2 n_B)^{(1/3)} \sim \frac{2 \pi}{L} \sqrt{l_{F,x}^2 + l_{F,y}^2 + l_{F,z}^2}
\end{equation}
with $F$ corresponding to the last plane wave filled by the atoms of each bath 
($l_{F,x}=2, l_{F_y}=l_{F,z}=0$ or combinations with two indexes equal to zero 
and one equal to 2).  With 33 particles per bath,  we have $k_F$ = 
0.04167$\mu_{13}$ for the first case (thermodynamic limit), and $k_F$ = 
0.04188$\mu_{13}$ when we use the last plane wave occupied in our discrete 
approximation.  When $N_1$ = $N_2$ = 33  the difference between both values is 
less than 1\%, explaining why this number of particles has been extensively 
used in DMC calculations of Fermi systems.   The energy corresponding to that 
last plane wave filled is:
\begin{equation}
\epsilon_F = \frac{\hbar^2 k_F^2}{2 m} = \frac{k_F^2}{4 \mu_{13}^2} \epsilon_0,
\end{equation}
from which the total energy of the two noninteracting baths, containing 66 particles in total, can be calculated in the thermodynamic limit:
\begin{equation}
E_F = \frac{3}{5} (N_1 + N_2) \epsilon_F.
\end{equation}
In our case, $E_F$ = 0.01719$\epsilon_0$, to be compared with 
the value obtained by summing up all the energies corresponding 
to the plane waves $\mathbf{k}_a=(2\pi/L)(l_{a,x},l_{a,y},l_{a,z})$,
up to the one corresponding to $k_F$,  $E_F$ = 0.01707$\epsilon_0$.  We will use that last value in the remainder of the paper. 


\section{Results}

\begin{figure}
\begin{center}
\includegraphics[width=0.8\linewidth]{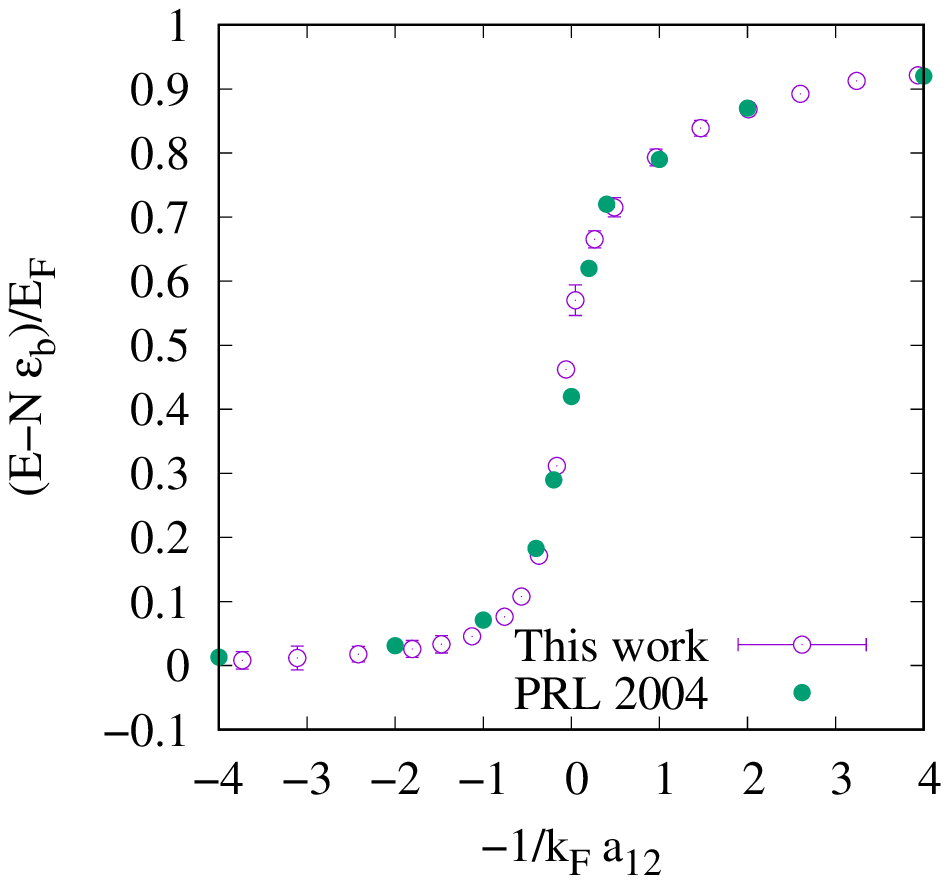}
\caption{Total energy of the system as a function of $-1/(k_F a_{12})$ ($a_{12}$ being the scattering length between the two-baths species)  formed by two baths and no impurity across the BEC-BCS crossover, obtained using the P\"oschl-Teller potential (open circles). Results obtained with a square-well potential in Ref. \onlinecite{jordibec} (PRL 2004, full circles) are included for comparison. Here, $E_F$ is the energy of two non-interacting Fermi baths. In the BEC branch, the total energy is corrected by subtracting the binding energy of all the $N$ pairings between atoms belonging to different baths, $N\epsilon_b$, which is zero in the BCS branch.}
\label{fig1}
\end{center}
\end{figure}

To test the accuracy of our results, with respect to those in previous 
works,  we compared  our simulation results for the system with two 
baths and no impurity with those of Ref.~\onlinecite{jordibec}, 
obtained using a square-well interparticle interatomic potential.  This comparison is 
displayed  in Fig. \ref{fig1} for interaction strengths in the range -4$<-1/(k_F 
a_{12}) <$ 4, $k_F$ being the Fermi wave vector.  For clarity,  we subtract from 
the total energy of the system comprising 66 atoms, $E$, the binding energy of 
$N$ pairs, $N\epsilon_b$, in the regime where those pairs can form, i.e., 
when $a_{12}>$0.  What we find is that the two sets of data agree within the 
statistical uncertainties of our calculations. 
In the unitary limit, there is a relationship between the total energy of
$2N$ atoms, $E$, and the energy of the two non-interacting Fermi baths,
$E_F$, given by $E=\xi E_F$.
We find $\xi=0.46(2)$, to be compared with $\xi=0.44(2)$ obtained in
Ref.~\onlinecite{panda} using the same potential, and $\xi=0.42(1)$
obtained for the square-well interaction in Ref.~\onlinecite{jordibec}.
These values are above the best available unbiased estimate,
$\xi\simeq0.37$, obtained using auxiliary-field quantum Monte Carlo,
which is free from the sign problem at unitarity~\cite{AFQMC}.

To gauge the accuracy of our result, finite-range effects must first be
taken into account. For a finite-range interaction, the value of $\xi$
depends on dimensionless parameters such as $r_e/r_0$ or, equivalently,
$k_F r_e$, where $r_e=2/\mu_{12}$ is the effective range of the
P\"oschl--Teller potential and
$r_0=[3/(8\pi n_B)]^{1/3}$. The universal zero-range limit corresponds to
$r_e/r_0\rightarrow 0$, or $k_F r_e\rightarrow 0$.
Using the same P\"oschl--Teller interaction as in Eq.  \ref{pt} , and for $2N$ = 66,  Gandolfi \textit{et al.}
~\cite{gandolfi} obtained $\xi=0.3923(4)$ for
$r_e/r_0\simeq0.02$, while for $r_e/r_0\simeq0.07$ they found
$\xi=0.4069(5)$. In our simulations, $r_e/r_0=0.0579$, for which we
obtain $\xi=0.46(2)$. A more systematic study of the effective-range
dependence was carried out by Forbes \textit{et al.}
~\cite{forbes}, also for $2N$ = 66, who parametrized the energy in terms of $k_F r_e$.
For our value, $k_F r_e\simeq0.1111$, their fit gives
$\xi\simeq0.4045$. These results suggest that a value around
$\xi\simeq0.40(1)$ is appropriate for a system with finite-range
corrections comparable to ours.

A further source of uncertainty is the nodal structure of the trial
wave function. Gandolfi \textit{et al.} ~\cite{gandolfi} 
showed 
that variational
optimization of a more flexible pairing wave function can lower the
fixed-node energy by approximately 4--7\%. 
An improvement of a comparable magnitude to our result would 
reduce the value of $\xi$ to the range 
$0.43(2)$--$0.44(2)$, bringing it considerably closer to the values expected
for comparable finite-range conditions. Our aim, however, is not to
provide a high-precision determination of $\xi$, but
to study the change produced by adding a single impurity to the two
Fermi baths both at unitary and away from it.
The quantities relevant to our conclusions are therefore mainly
energy differences between the system containing the impurity and the
corresponding impurity-free system. In both calculations the same
fermionic nodal structure is used, while the impurity-bath correlation
factors are nodeless. We therefore expect a substantial cancellation
of the fixed-node bias in these energy differences. This cancellation
need not be exact, since the optimal nodal surface could itself be
modified by the presence of the impurity, but the absolute fixed-node
bias of the bath energy should affect the corresponding energy
differences much less strongly. The agreement with the results of
Ref.~\onlinecite{jordibec} away from unitarity, shown in
Fig.~\ref{fig1}, provides additional support for this interpretation.

 
\begin{figure}
\begin{center}
\includegraphics[width=0.8\linewidth]{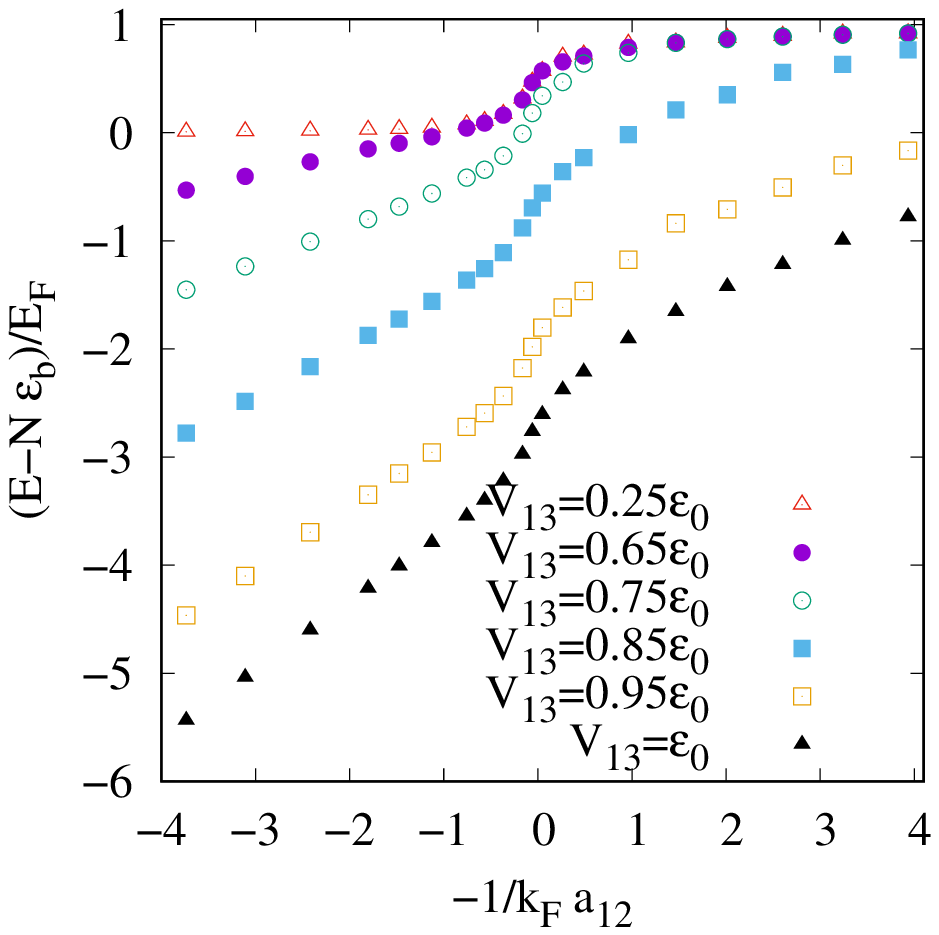}
\caption{
Same as in Fig. \ref{fig1}, but for a system including an impurity interacting with the particles in both baths with different strengths, or equivalently with different negative scattering lengths. The error bars are of the size of the symbols and not displayed for simplicity.  
The corresponding scattering lengths are (from lowest to highest displayed $V_{13}$ values) $a_{13}$= -0.54$\mu_{13}^{-1}$,  -2.89$\mu_{13}^{-1}$, -4.62$\mu_{13}^{-1}$, -8.64$\mu_{13}^{-1}$,-28.66$\mu_{13}^{-1}$, -$\infty$, respectively. 
}
\label{fig2}
\end{center}
\end{figure}

In Fig.~\ref{fig2}, we show our FN-DMC results for the full system,
including the two Fermi baths and the impurity, for different impurity--atom 
interaction strengths. In the BEC regime of the baths, the data are presented 
after subtracting the binding energy of the same $N$ pairs considered in Fig. 
\ref{fig1} from the total energy of the system, $N\epsilon_b$.  The resulting energies are 
expressed in units of $E_F$. Their  behavior differs significantly 
from that shown in Fig.~\ref{fig1}.  Except at very weak impurity--atom 
interactions, the energy decreases noticeably both in  the BCS and on the BEC 
regimes of the two baths. 

\begin{figure}
\begin{center}
\includegraphics[width=0.8\linewidth]{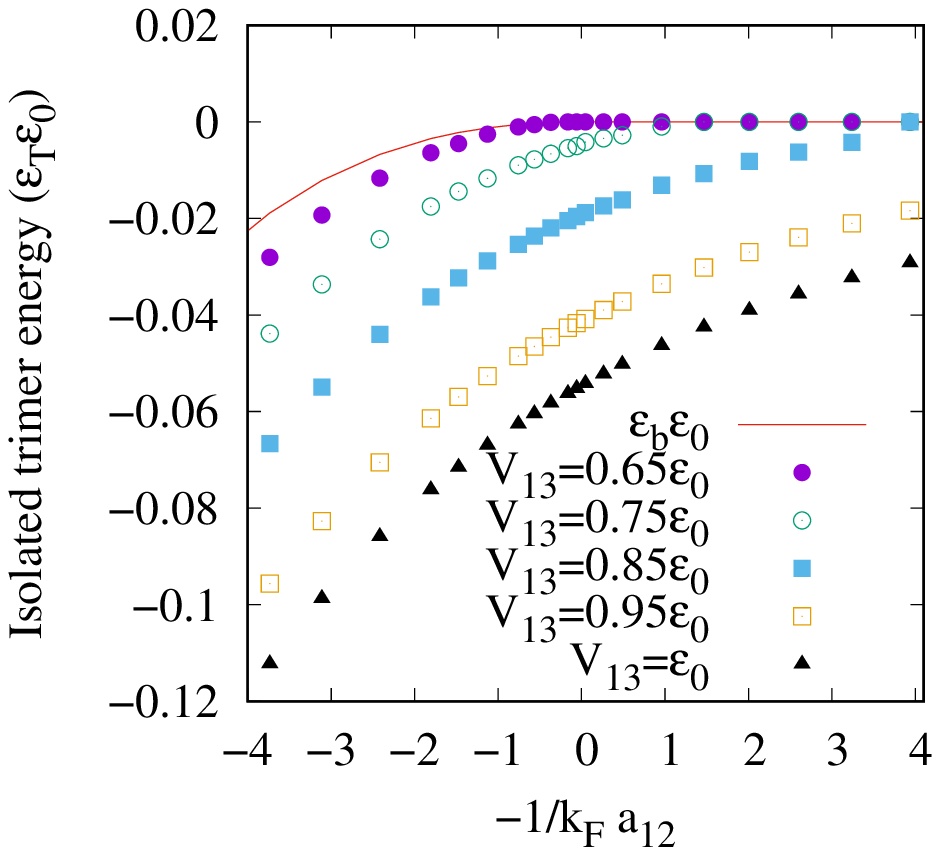}
\caption{
Energy of a system containing three isolated atoms, $\epsilon_T$, in units of $\epsilon_0$.  The error bars are smaller than the size of the symbols.The solid line corresponds to the two-body binding energy obtained from Eq. \ref{binding}.
}
\label{fig3}
\end{center}
\end{figure}

The origin of this effect must be the 
association of the impurity with two atoms, one belonging to each bath.   To check this interpretation, we first calculated the bound-state energy, $\epsilon_T$, of three isolated atoms interacting through the same potentials as in the impurity--two-bath system. $-\epsilon_T$ (a positive magnitude), can be thought as a separation energy, i.e., the energy we have to give to the three atoms to break up the trimer.
To solve  
the three-atom Schrodinger equation, we considered the same $\phi(r)$ and 
$\varphi(r)$ functions defined above as trial wave functions.  The results are 
shown in Fig.~\ref{fig3}. In the same figure, we also show the interaction  energy ($\epsilon_b \epsilon_0$) of 
a single pair of atoms, obtained from Eq.~\ref{binding}. This latter quantity 
can also be interpreted as the energy of an impurity located far from a bound 
pair of atoms.  Since the values for the $V_{13}$= 0.50$\epsilon_0$ case are on top of those corresponding to $\epsilon_b$, we can say that the impurity does not bind to any pair of bath atoms.  
We find that the binding energy of those trimers decreases significantly from the one corresponding to a dimer plus an isolated atom for
appropriate values of the scattering lengths.  
Interestingly, we can have 
bound trimers when no dimer is present, i.e.,  when $a_{12}$ and $a_{13}$ are 
both negative,  as shown in Fig. \ref{fig3} for $V_{13}>0.75\epsilon_0$.  Those three-atom 
sets are called Borromean trimers~\cite{borromean, jordi2014}.  On the other 
hand,  when the $a_{12}$ changes from negative to positive and a $1,2$ bound 
dimer is present, we have tango  trimers \cite{borromean,jordi2014,tango}.  

\begin{figure}
\begin{center}
\includegraphics[width=0.8\linewidth]{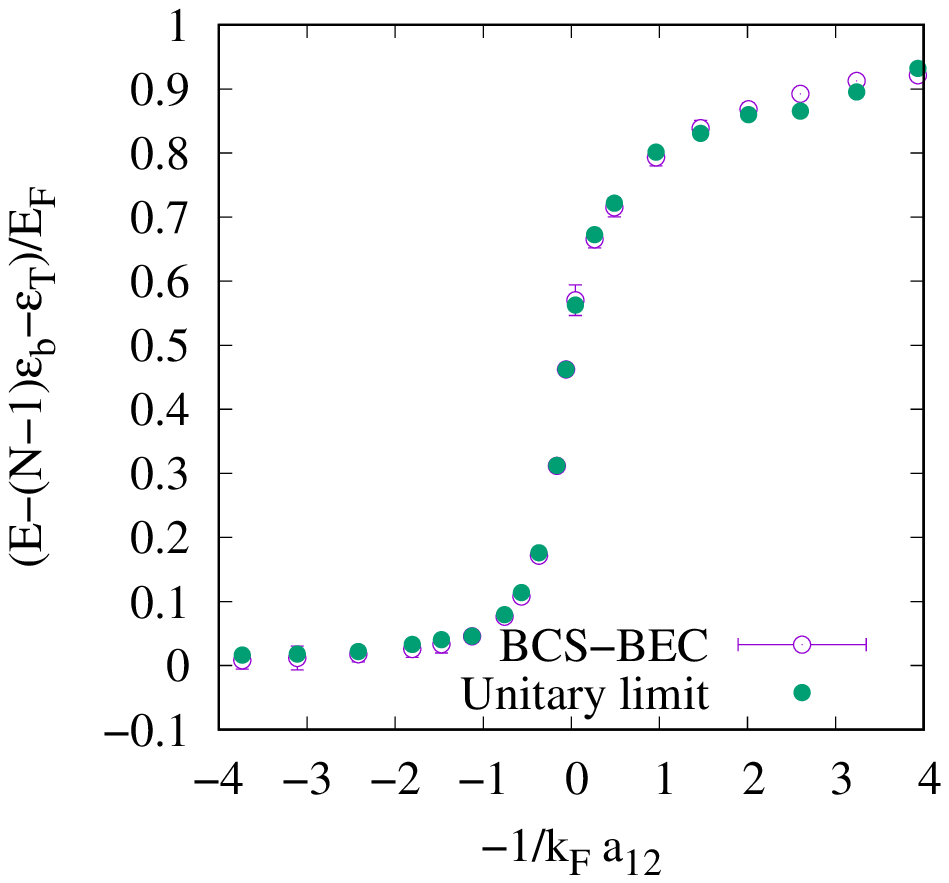}
\caption{
Energy of the system including the impurity and the two baths, after subtracting the binding energy of  $N$-1 $(1,2)$ pairs and one trimer in the impurity-bath unitary limit (full circles). The energy of the two baths without impurity, under the same conditions as in Fig. \ref{fig1}, is also shown for comparison (open circles). The error bars are of similar size in both cases and, for clarity, are displayed only for the latter. 
}
\label{fig4}
\end{center}
\end{figure}

Once the different values of $\epsilon_T$ had been obtained, we calculated
$E-(N-1)\epsilon_b-\epsilon_T$, in units of $E_F$, and display the
results in Fig.~\ref{fig4}. The energy of the corresponding impurity-free
two-bath system, already shown in Fig.~\ref{fig1}, is included for comparison.
We show only the case of the impurity in the unitary limit, since it is 
representative of all the other cases. 
We can see that, within the statistical uncertainties, the two curves are nearly identical.
This result supports a simple cluster interpretation. In the BEC regime, the
system containing the impurity can be viewed as $N$ clusters, namely $N-1$
dimers and one trimer. After subtracting their internal binding energies, the
remaining contribution is essentially the same as the one obtained for the $N$
dimers of the impurity-free system. Thus, at the very low density considered
here, replacing one dimer by one trimer does not produce a detectable change
in the residual cluster--cluster interaction within our statistical resolution.
In addition, since there is only one trimer in the simulation cell, its composite fermionic character does not lead to 
an additional Fermi-statistical
contribution.

\begin{figure}
\begin{center}
\includegraphics[width=0.8\linewidth]{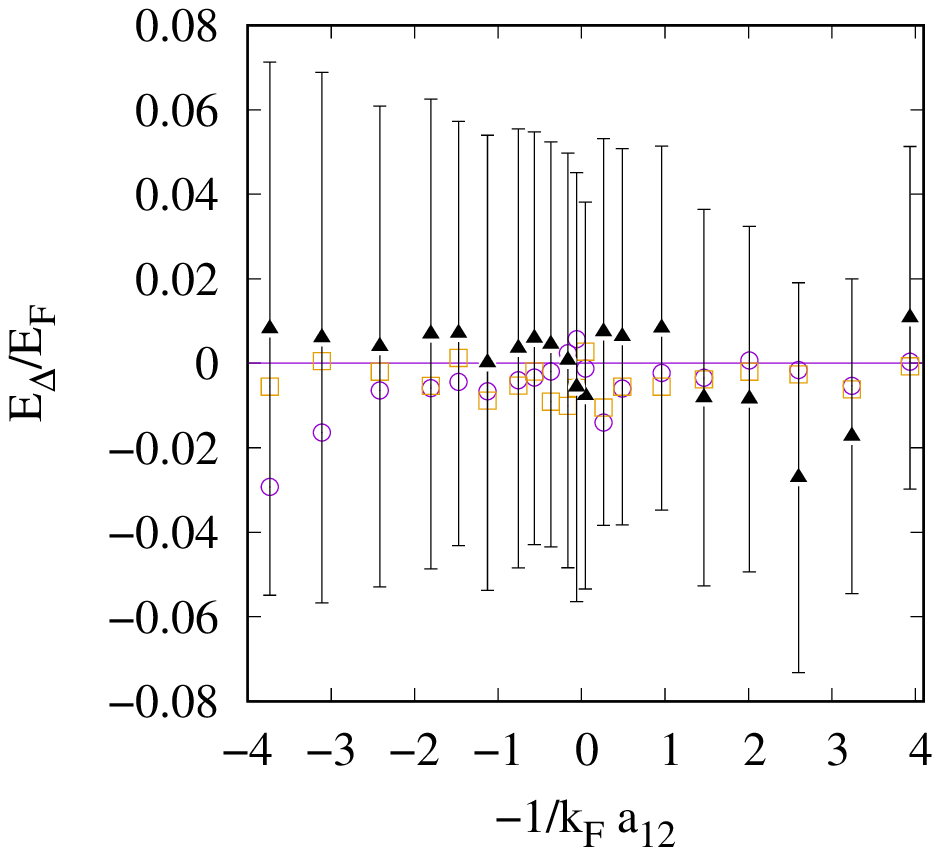}
\caption{
Residual energy, in units of $E_F$ for 
$V_{13}=0.50\epsilon_0$, open circles;
$V_{13}=0.65\epsilon_0$, open squares, and
$V_{13}=\epsilon_0$, solid triangles.
Error bars are shown only for the unitary case, for which they are the
largest. In all cases, the residual energies are compatible with zero,
indicating negligible interactions between the resulting clusters
within our statistical uncertainty.
}
\label{fig5}
\end{center}
\end{figure}

To examine this interpretation more directly, in Fig.~\ref{fig5} we plot the
residual energy
\begin{equation}
E_{\Delta}=E-N \epsilon_b - E_{\rm BCS-BEC}-\Delta\epsilon ,
\label{residuo}
\end{equation}
in units of $E_F$, where
\begin{equation}
\Delta\epsilon=\epsilon_T-\epsilon_b.
\end{equation}
is the difference in energy of the $N+1$ and $N$ systems, that we calculated from an isolated trimer and a bound dimer using Eq.  \ref{binding}, i.e. 
neglecting 
dynamical correlations between particles.
In Eq. (\ref{residuo}), $E_{\rm BCS-BEC}$ is the energy of the two baths 
without the impurity after being corrected by $N\epsilon_b$, i.e., the one 
displayed in Fig. \ref{fig1}.  When no trimer is bound in vacuum, 
$\epsilon_T=\epsilon_b$
(with $\epsilon_b=0$ on the BCS side), and therefore
$\Delta\epsilon=0$. In this case, $E_{\Delta}$ measures the
interaction of the impurity with both baths. When a bound trimer exists,
$E_{\Delta}$  measures instead the difference between the 
interaction energy of the $N$ clusters in the system
($N-1$ dimers and one trimer) and that of the $N$ dimers in the
corresponding impurity-free system.

For $V_{13}=0.50\epsilon_0$ (open circles), for which
$\epsilon_T=0$, the impurity--medium interaction
appears to be slightly attractive. For $V_{13}=0.65\epsilon_0$
(open squares) and $V_{13}=\epsilon_0$ (solid triangles), for which
isolated bound trimers are present, the residual energy is also close
to zero. In all three cases, however, the statistical uncertainties
are large enough for $E_{\Delta}$ to be compatible with zero.
The same behavior is found for all the impurity--bath interaction
strengths considered in this work, whose results are not shown for
clarity. Error bars are displayed only for the unitary case, for which
they are the largest, approximately twice as large as those of the other two cases.

That $E_{\Delta} \sim 0$ implies that when no bound trimer exists, the present statistical accuracy
does not allow us to determine whether the impurity forms a polaron-like
state or behaves essentially as a non-interacting impurity. In the same way,  when a bound
trimer exists in vacuum, the results indicate that,
within our statistical resolution, replacing one bath dimer by a trimer
does not produce a detectable change in the interaction energy
of the system after subtraction of the corresponding binding energies of the $N$ clusters.
Finally, our results show that the difference in energies between the system 
with and without impurity is, to a large extent, the same 
quantity determined by the few body 
quantities $\epsilon_T$ and $\epsilon_b$.

\begin{figure}
\begin{center}
\includegraphics[width=0.8\linewidth]{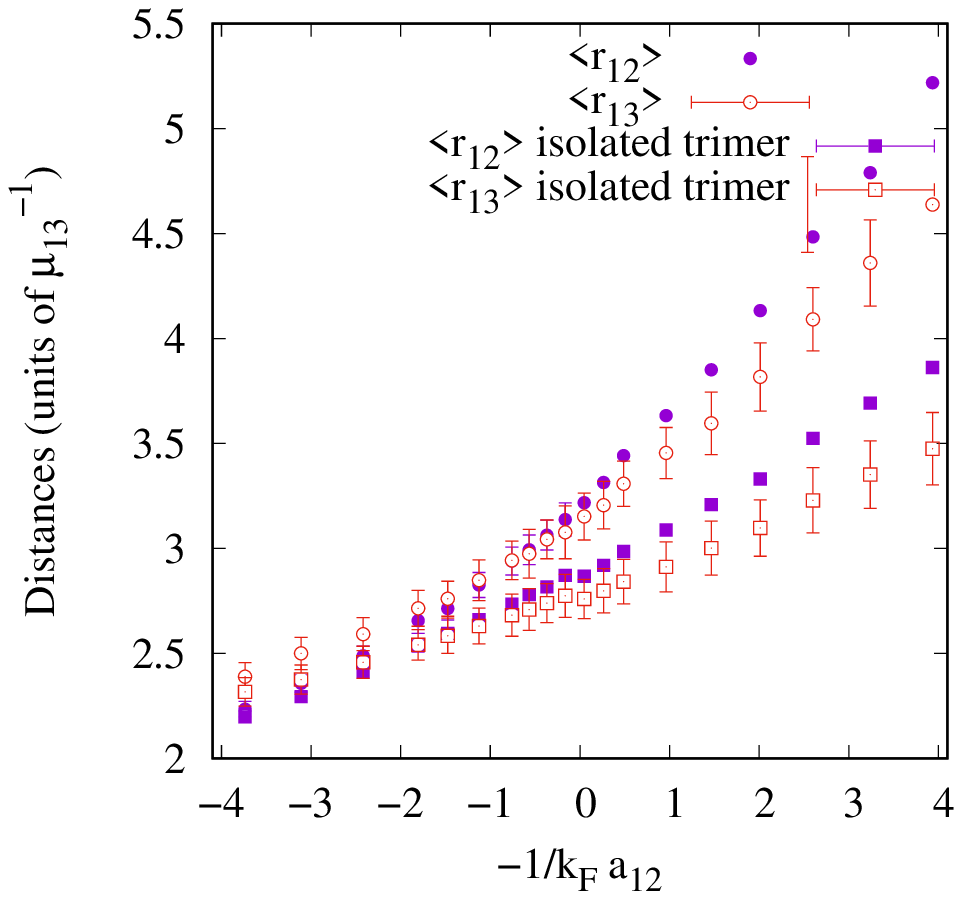}
\caption{
Average interparticle distances in the isolated trimer (squares) and in the trimer embedded in the two Fermi baths (circles), in the impurity-bath unitary limit.  When not shown, the error bars are comparable to the ones displayed.
}
\label{fig6}
\end{center}
\end{figure}

The difference between the geometry of trimers embedded in the baths and that of 
isolated 
three-atom clusters is shown in Fig. \ref{fig6}.  To avoid any bias related to the trial function, pure estimators were used in both cases.  We plot the average distance 
between the two bath atoms associated with the impurity,  $\langle r_{12} 
\rangle$, together with the mean impurity-bath distance,  $\langle r_{13} 
\rangle$ = $\langle r_{23} \rangle$.  Those last distances were averages of the closest impurity-bath distances calculated at each Monte Carlo step.
We chose this observable rather than calculating the mean distance from the impurity to atoms within a given cutoff in order to ensure that only one atom from each bath is considered. To allow the atoms associated with the impurity to change during the simulation, these distances were recalculated at every Monte Carlo step.  Thus, the means obtained correspond to the entire distribution of closest distances and are not simply the smallest of each simulation run. 
As in the previous discussion, we focus on the unitary limit of the 
impurity-bath interaction.
The overall size of the trimer is found to be smaller in the isolated case with respect to the embedded cluster for the same parameters.  However, this difference decreases considerably in the BEC regime.  This enlargement is likely related to the fact that the atoms associated with the impurity must still satisfy the Fermi statistics imposed by their respective baths. 

\begin{figure}
\begin{center}
\includegraphics[width=0.8\linewidth]{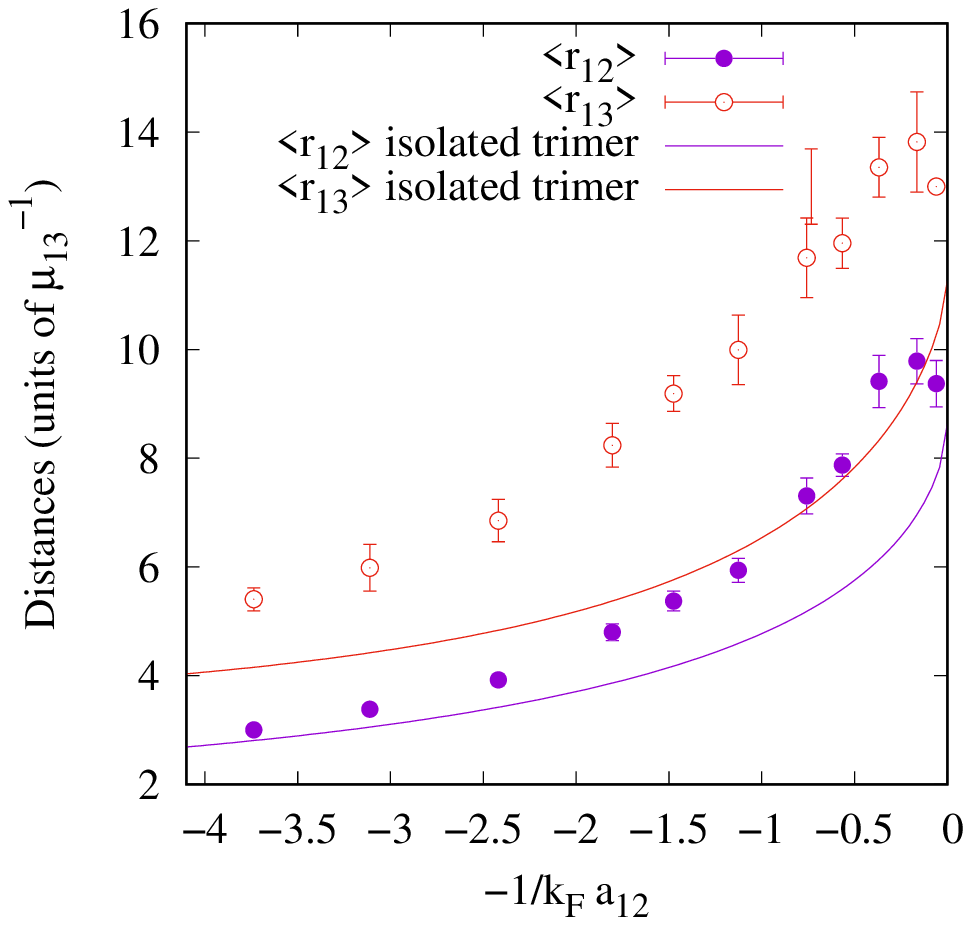}
\caption{
Same as in the previous figure for $V_{13}$=0.75 but only in the BEC regime.  The solid lines are smooth fits to the isolated-trimer data.
}
\label{fig6b}
\end{center}
\end{figure}

That trend in the sizes of the clusters inside and outside the two-bath environment in the BEC regime, beyond the unitary limit, can be seen in Fig. \ref{fig6b}, which shows the case $V_{13}$=0.75$\epsilon_0$ ($a_{13}$=-4.62$\mu_{13}^{-1}$). The only difference with respect to the previous figure is that the trimers are much larger than in the case of the stronger impurity-bath interaction corresponding to $a_{13}\rightarrow -\infty$.  
To provide information about the geometry of those clusters, 
we calculate for $-1/(k_F a_{12})$=-3.74 (leftmost value in Figs. \ref{fig6} and \ref{fig6b}), the ratio
$\mathcal{R}=\langle r_{13}\rangle/ \langle r_{12}\rangle$, that
decreases as the impurity-bath interaction becomes stronger, as can be seen in Table \ref{table0}.  
Thus, the weakly bound trimer is initially a highly elongated object, but as the impurity-baths attraction increases, the three particles evolve towards a more compact configuration.  
This suggests that there is an evolution between an independent impurity,  far away from a dimer and a full trimer. 
This trimer disappears for  $V_{13} \leq$ 0.50$\epsilon_0$, 
in agreement with the energy results of Fig. \ref{fig3} for isolated 
clusters. 
\begin{table}[t]
\caption{Ratio $\mathcal{R}=\langle r_{13}\rangle/\langle r_{12}\rangle$
for different impurity--bath interaction strengths.}
\label{table0}
\begin{ruledtabular}
\begin{tabular}{ccc}
$V_{13}$ (units of $\epsilon_0$) & $a_{13}$ (units of $\mu_{13}^{-1}$) &$\mathcal{R}$ \\
\hline
0.55 & -1.92 & 5.5 \\
0.60 & -2.34 & 4.0 \\
0.65 & -2.89 & 3.0 \\
0.75 & -4.62 & 1.8 \\
1.00 & -$\infty$ & 1.1 \\
\end{tabular}
\end{ruledtabular}
\end{table}

\begin{figure}
\begin{center}
\includegraphics[width=0.8\linewidth]{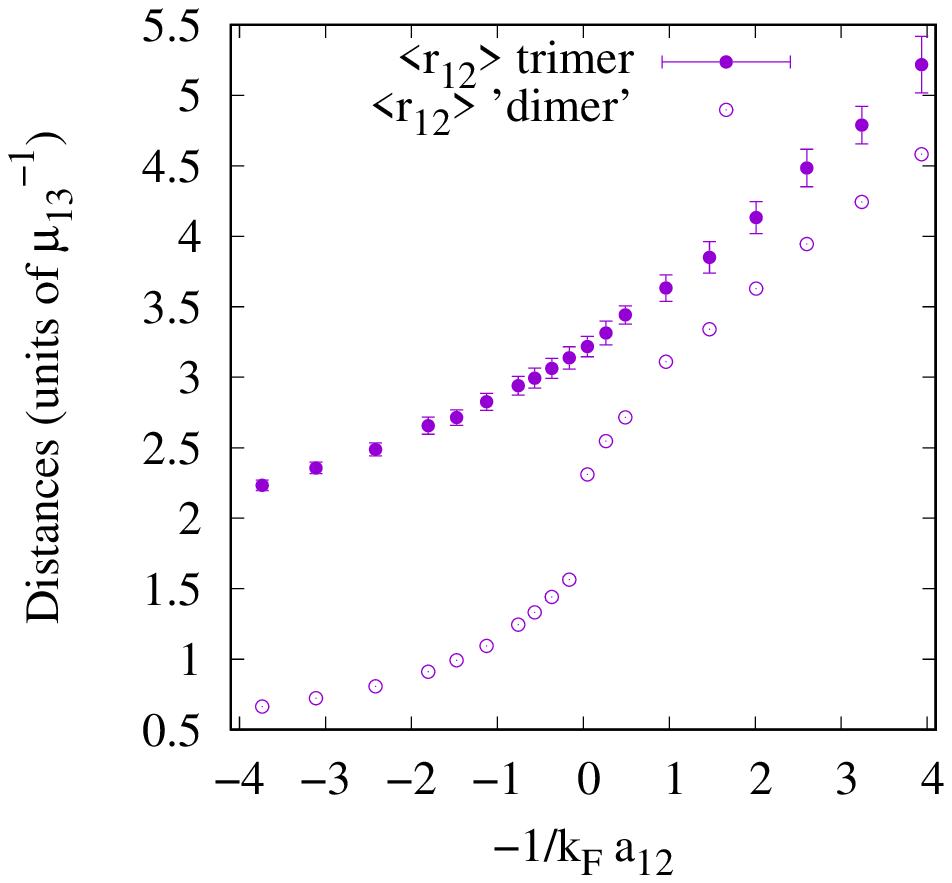}
\caption{
Average \(1\text{--}2\) distance between the two bath atoms associated with the impurity and forming the embedded trimer (full symbols), compared with the average minimum distance between atoms belonging to different baths (open symbols). For each configuration, the latter quantity is obtained by selecting the shortest distance among all possible \(1\text{--}2\) pairs,  either they are bound to form a dimer or not. The results correspond to the impurity-bath unitary limit.
}
\label{fig7}
\end{center}
\end{figure}

The relation between the geometry of the embedded trimer and that of the 
bath-bath pairs can be further analyzed from Fig. \ref{fig7}, again for the 
unitary limit of the impurity-bath interaction. The full symbols correspond to 
the same $\langle r_{12} \rangle$ distance displayed in Fig. \ref{fig6}, namely 
the average separation between the two bath atoms associated with the impurity. 
The open symbols represent instead the average minimum distance between atoms 
belonging to different baths. More precisely, for each many-body configuration 
we computed all possible (1\text{--}2) distances, selected the shortest one, and 
then averaged this quantity over the Monte Carlo sample. In the BEC regime, this 
 minimum distance provides an estimate of the typical separation 
between the two atoms of a dimer.  
On the other hand,  in the BCS
regime, where no bound dimers exist, it simply represents the minimum
distance between atoms belonging to different baths and is used as a
reference for comparison with the corresponding separation within the
trimer. 
In the BEC regime,  $\langle r_{12} \rangle$ for a dimer is at least a factor of two smaller than 
the 
separation between the two bath atoms forming the trimer, two being the ratio in the unitary limit of the two baths. Therefore,  the trimer 
cannot be interpreted as simply a dimer with an impurity attached to it. 

A similar conclusion holds on the BCS side of the crossover. Even when 
no dimer exists, the distance between the two bath atoms associated with the 
impurity remains larger than the typical closest (1\text{--}2) distance in the 
many-body configuration. We used the closest distance instead of the distance to atoms up to a particular cutoff to ensure that we consider a single pair, and to calculate that parameter in the same way in both cases to make the comparison as fair as possible.  This indicates that the trimer is not simply built by 
the impurity attaching itself to the two closest available pair of bath atoms. 
Rather, within the range of interactions considered here, the impurity induces a 
well-defined three-body structure embedded in the two Fermi baths.

\begin{table*}[t]
\caption{
Energies for different values of $\mu_{13}'$ in 
the unitary  limit for
different values of 
bath-bath interaction strengths, -$1/(k_Fa_{12})$.
$E$ is the total energy of the system containing the impurity, while 
the residual energy is defined as in Eq. \ref{residuo}. 
All energies are expressed in the common reference unit \(\epsilon_0=\hbar^2\mu_{13}^2/m_r\), where \(\mu_{13}\) denotes its reference value, and it is the same for all cases.   Errors correspond to the figure in parenthesis. 
}
\label{tabla1}

\begin{ruledtabular}
\begin{tabular}{ccrrrr}
$\mu_{13}'$ &
-$1/(k_Fa_{12})$ &
$E -N\epsilon_b$ &
$E_{\rm BCS-BEC}$ &
$\Delta \epsilon$ &
$E_{\Delta}$ \\
\hline
0.5$\mu_{13}$ & $3.96$ & 6.8(1) 10$^{-3}$ & 1.58(3) 10$^{-2}$ & -8.7(1) 10$^{-3}$ & -3(3) 10$^{-4}$ \\
    & $0.0$  & -1.36(3) $10^{-2}$  & 8(1) 10$^{-3}$ & -2.15(2) 10$^{-2}$ & -1(4) 10$^{-4}$  \\
    & $-4.33$  & -4.17(3) 10$^{-2}$ & 1(4) 10$^{-4}$ & -4.17(3) 10$^{-2}$ & 0(7) 10$^{-4}$  \\
\hline
 $\mu_{13}$ & $3.96$ &-1.33(5) 10$^{-2}$ & 1.58(3) 10$^{-2}$ & -2.92(5) 10$^{-2}$ & -1(7) 10$^{-4}$ \\
    & $0.0$  & -4.73(6) 10$^{-2}$ & 8(1) 10$^{-3}$ & -5.52(6) 10$^{-2}$ & -1(8)  10$^{-4}$  \\
    & $-4.33$  & -9.9(1) 10$^{-2}$ & 1(4) 10$^{-4}$& -9.9(1) 10$^{-2}$ &   0(1) 10$^{-3}$\\
\hline
1.5$\mu_{13}$ & $3.96 $ & -3.38(1) 10$^{-2}$ &1.58(3) 10$^{-2}$  &  -4.88(1) 10$^{-2}$ & 2(3) 10$^{-4}$ \\
    & $0.0$  & -7.7(1) 10$^{-2}$ & 8(1) 10$^{-3}$ & -8.5(1) 10$^{-2}$ & 0(1) 10$^{-4}$ \\
    & $-4.33$  & -0.148(1) & 1(4) 10$^{-4}$ & -0.148(1) & 0(1) 10$^{-3}$ \\
\end{tabular}
\end{ruledtabular}
\end{table*}
All the results discussed above were obtained for a fixed ratio between
$\mu_{12}$ and $\mu_{13}$. To test the robustness of our conclusions
against changes in this ratio, we performed additional calculations at
different $\mu_{12}/\mu_{13}$ ratios while we kept constant the $a_{13}$ scattering lengths by changing the $V_{13}$ parameter in Eq. \ref{pt2} accordingly.  An example of what happens can be seen in Table \ref{tabla1}
for the unitary limit of the 
the impurity--bath interaction when we take
$\mu_{13}'=0.5\mu_{13}$ and $\mu_{13}'=1.5\mu_{13}$ while keeping
$\mu_{12}$ fixed. For each value of $\mu_{13}'$, we considered three
different bath--bath interaction strengths, characterized by
$-1/(k_Fa_{12})$. 
In that Table, we can see that there is a huge variation both in the corrected total energy of the systems, $E -N\epsilon_b$, and in the difference $\Delta \epsilon$ = $\epsilon_T-\epsilon_b$ (when $-1/(k_F a_{12}) \geq 0$, $\epsilon_b$ = 0) when we change the values of $\mu_{13}'$.   However,  those changes compensate each other when we calculated $E_{\Delta}$ for each system.   
For all the values of $\mu_{13}'$ and $a_{12}$ considered,
$E_{\Delta}$ is compatible with zero within the statistical
uncertainties. Thus, although the absolute energies and the trimer
binding energies depend appreciably on the relative interaction ranges,
replacing one bath pair by a trimer does not
produce a detectable change in the residual interaction energy between the $N$ clusters in the range of $\mu_{13}'$'s considered. The same happens when instead of a trimer we have an isolated impurity. 

\section{Conclusions}

By means of FN-DMC calculations, we have studied the behavior of a single 
impurity embedded
in two Fermi baths across the BCS-BEC crossover of the bath-bath interaction. 
The atoms in the two baths and the impurity were considered to have the same 
mass. The impurity interacted attractively and symmetrically with all other 
atoms, but in no case those interaction strengths were  deep enough to produce 
impurity-atom bound pairs. This setup can be regarded as a simplified model for 
a mixture of the three lowest hyperfine states of $^6$Li, a system with three 
fermionic components of equal mass whose interactions can be tuned by using a 
magnetic field \cite{li6-0,li6-1,li6-2}. However,  even though the scattering lengths of those interactions cannot be varied independently of each other, we do have particular values of the magnetic fields for which $a_{12} >$0  and $a_{13}=a_{23} < 0$ \cite{li6-1} or $a_{12} >$0  and $a_{13} \neq a_{23} < 0$ \cite{li6-2}, whose results can be compared to our calculations. 

In our calculations, we have paid particular attention to the range of 
impurity-bath interactions in which trimer formation is observed. 
 In fact, for $V_{13} \leq$0.50$\epsilon_0$  the impurity behaves approximately as a 
non-interacting particle inside those baths. Although the energy difference 
between the impurity + two-baths system and the impurity-free setup appears to 
be slightly negative, the size of the error bars makes this attraction  
compatible with zero. The absence of a trimer in this regime is also confirmed 
by making a calculation including only three atoms,  and observing that the 
energy obtained for the $V_{13}$=0.50$\epsilon_0$ case lies on top of the binding-energy 
curve of the isolated dimer. 

When the impurity--bath attraction is sufficiently strong, the impurity binds 
one 
atom from each bath and forms a three-body cluster. On the BCS side, where none 
of the two-body subsystems is bound, these states are Borromean-like. As the 
bath-bath interaction is increased across the crossover, and eventually 
supports a dimer, the corresponding three-body states evolve  into 
tango-like trimers. In the BEC regime, subtracting the energy of one trimer and
of the remaining bath-bath dimers from the total energy, one can see 
the equation of state of the impurity-free system is restored.
 This supports 
a dilute-cluster picture in which the replacement of one bath dimer by a trimer 
does not produce a measurable change in the residual cluster-cluster 
contribution at the density considered here.  This conclusion is robust with respect to reasonable changes in the $\mu_{12}/\mu_{13}$ ratio. 

The analysis of the interparticle distances further shows that the 
embedded clusters are genuine three-body structures. In particular, the 
separation between the two bath atoms associated with the impurity is 
substantially larger than the typical separation between atoms forming bath 
dimers in the BEC regime. Likewise, on the BCS side, the trimer is not simply 
formed by attaching the impurity to the closest available pair of bath atoms. 
Although the embedded trimers are noticeably larger than their isolated 
counterparts, especially in the BCS regime, their overall binding energies and geometries 
remain similar. Our results therefore provide a microscopic description of 
impurity-induced trimer
formation in two interacting Fermi baths across the BCS--BEC crossover, 
and they are consistent with the experimental identification of those three-atom 
structures \cite{Braaten2010, Nakajima2010, Science}. 
The possible connection of these states with Efimov physics remains open, 
since establishing such a character would require resolving an excited-state 
spectrum and identifying the associated discrete scaling pattern, something 
very difficult within the FN-DMC method.

\begin{acknowledgments}
We acknowledge financial support from Ministerio Espa\~nol de Ciencia e Innovaci\'on under grant No.  PID2023-147469NB-C21 and Universidad Pablo de Olavide 
group GrIN-UPO FQM-205.
The authors acknowledge, too, the use of the computer facilities of C3UPO at the Universidad Pablo de Olavide, de Sevilla via
"Plan Propio de Investigación y Transferencia (2023-2026) de la UPO, por la Consejería de Universidades, Investigación e Innovación de la Junta de Andalucía y por el Programa FEDER Andalucía 2021-2027, 2023/00002/014."
\end{acknowledgments}

\bibliography{impureza}

\end{document}